\documentclass[
aps,
prb,
floatfix,
longbibliography,
reprint
]{revtex4-2}

\usepackage[T1]{fontenc}

\usepackage{amsmath}
\usepackage{amssymb}
\usepackage{amsfonts}
\usepackage{mathptmx}
\usepackage{graphicx}
\usepackage{color}
\usepackage[dvipsnames]{xcolor}
\usepackage{bbm}
\usepackage[breaklinks=true,colorlinks=true,linkcolor=blue,urlcolor=blue,citecolor=blue]{hyperref}
\usepackage{array}
\usepackage{booktabs}
\usepackage{placeins}

\DeclareMathAlphabet{\mathcal}{OMS}{cmsy}{m}{n}

\begin{document}

\title{Geometry-Induced Current Decrowding in Superconducting 3D Constrictions}

\author{Igor Bogush\textsuperscript{1,2}}
\email{igori.bogus@tu-braunschweig.de}
\author{Oleksandr Dobrovolskiy\textsuperscript{1,2,3}}

\affiliation{
\textsuperscript{1}Cryogenic Quantum Electronics, Institute for Electrical Measurement Science and Fundamental Electrical Engineering, Technische Universit\"at Braunschweig, Hans-Sommer-Str. 66, D-38106 Braunschweig, Germany\\
\textsuperscript{2}Laboratory for Emerging Nanometrology (LENA), Technische Universit\"at Braunschweig, Langer Kamp 6A-B, D-38106 Braunschweig,  Germany\\
\textsuperscript{3}FLUXONICS---The European Foundry for Superconducting Electronics e.V., 38116 Braunschweig, Germany}

\begin{abstract}
Current crowding is a ubiquitous limitation in nanoscale devices, where confinement and sharp features distort current flow, generating localized current-density hotspots and premature failure. In superconductors, this effect suppresses the order parameter and promotes vortex nucleation, reducing the critical current below the intrinsic depairing limit.
Here, we demonstrate that 3D shaping of superconducting nanoarchitectures overcomes geometric current crowding through spatial current redistribution, establishing a geometry-induced decrowding effect. Using finite-element modeling based on the time-dependent Ginzburg--Landau equation, we reveal that 3D constrictions exhibit a pronounced and tunable response to moderate in-plane magnetic fields---a functionality absent in planar geometries. This field-controlled geometry enables critical-current modulation and unveils new vortex-dynamics regimes, including a non-reciprocal critical current. Furthermore, curvature and finite thickness fundamentally alter vortex nucleation by enabling apex-mediated entry of single vortex lines, followed by their 3D splitting into two vortex filaments---a mechanism that does not occur in 2D manifolds.
Our findings demonstrate 3D geometric engineering as a design paradigm for superconducting nanoarchitectures, offering control over current distribution and vortex dynamics in devices approaching the depairing limit.
\end{abstract}

\maketitle
\section{Introduction}
\vspace{-3mm}
Superconducting currents carry electrical charge without dissipation, yet this state is inherently fragile: when the local current density exceeds the critical value, superconductivity is suppressed and a resistive state emerges. In geometrically nonuniform superconducting structures, such as those containing sharp corners~\cite{Mcc16nal}, edge defects~\cite{Dob20nac}, or constrictions~\cite{Emb17nac}, geometric constraints on current flow lead to a local enhancement of the current density known as current crowding~\cite{Cle11prb,Ada13apl}. The resulting local current density can approach or exceed the depairing limit even when the total transport current remains well below its nominal critical value, thereby suppressing the experimentally accessible critical current of the device \cite{hor12apl,hen12prb,jon22pra}. So far, mitigating current crowding in planar superconducting nanostructures has primarily relied on geometric optimization, such as corner rounding and constriction tapering \cite{Cle11prb}. These approaches are intrinsically limited by the two-dimensional (2D) geometry.

The emergence of three-dimensional (3D) superconducting nanoarchitectures~\cite{Mak22adm,Fom22apl} has opened access to functionalities unavailable in planar geometries. Advances in 3D nanofabrication, including strain-relaxation-driven self-rolling~\cite{Thurmer08,Thurmer10} and direct-write deposition by focused electron and ion beams (FEBID/FIBID)~\cite{cor19nan,por19acs,Fer20mat,Hof23arx}, have enabled superconducting structures with enhanced microwave response~\cite{Loe19acs}, discrete resistance transitions~\cite{cor19nan}, and geometry-dependent magnetic behavior~\cite{zha26afm}. While local thickness variations have been used to alleviate current crowding in planar superconducting nanowire-based detectors \cite{Bag21sst,Xio22sst}, current redistribution arising intrinsically from the 3D geometry of a curved superconducting membrane has remained largely unexplored.

Here, we demonstrate that extending superconducting constrictions into the third dimension introduces a fundamentally different mechanism of current redistribution, which we term the geometry-induced current decrowding effect. Distinct from approaches based on local thickness engineering \cite{Bag21sst,Xio22sst}, we exploit the 3D cross-sectional geometry to maintain an effective current-carrying area comparable to that of the unconstricted leads. This redistributes the current through the third dimension and thereby suppresses the current crowding characteristic of planar constrictions with the same in-plane footprint. 

To uncover the physical origin of geometry-induced current decrowding, we combine 2D and 3D finite-element simulations of the time-dependent Ginzburg--Landau (TDGL) equation, which has recently been adapted to curved 3D superconducting nanoarchitectures \cite{Fom12nal,cor19nan,bog24prb,dee25acs,dee26acs,Mem25nah,bog25cpc}. We show that 3D geometry not only reduces current crowding but also reshapes vortex nucleation. Namely, curvature modifies the screening-current topology, enabling critical-current control by moderate in-plane magnetic fields and driving a transition from edge-mediated to apex-mediated vortex entry. This geometry-induced change in nucleation mechanism breaks the symmetry between opposite current polarities and produces a non-reciprocal critical current, revealing a route toward geometry-engineered superconducting diode behavior \cite{lyu21nat,mol25nat,Dob20pra}. Furthermore, 3D geometry fundamentally alters the coupling of superconducting transport to the orientation of the applied magnetic field. This geometry-enabled field sensitivity provides an additional degree of freedom for spatially selective control of the critical current and vortex dynamics.

Our findings establish a geometry-induced current decrowding effect as a design principle for superconducting nanostructures. Using the third dimension to tailor current flow and vortex dynamics provides a strategy to overcome geometric limitations and drive superconducting devices toward their intrinsic performance limits, with potential applications ranging from single-photon detectors \cite{Bag21sst} to Josephson weak links \cite{Yus23nal}.

\section{Model}

\begin{figure}[t!]
    \centering
    \includegraphics[width=8.5cm]{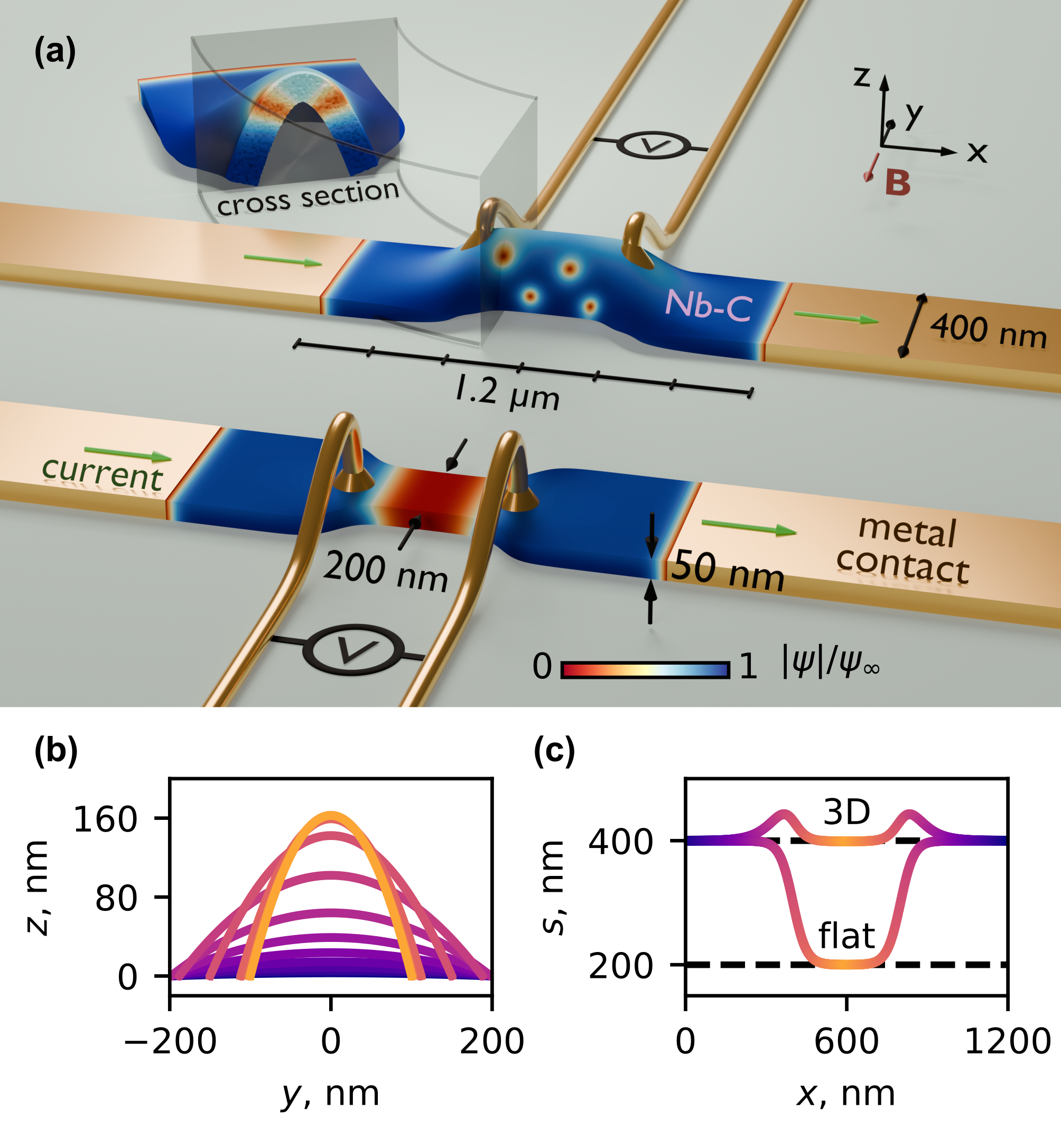}
    \caption{\textbf{(a)} Sketches of the 3D curved (top) and planar (bottom) superconducting constrictions, overlaid with the simulated spatial distributions of the superconducting order parameter. Arrows indicate the directions of the transport current and the applied in-plane magnetic field $\mathbf{B}$. \textbf{(b)} Transverse height profiles $z(y)$ of the 3D constriction for selected longitudinal positions $x$, indicated by the corresponding colors in \textbf{(c)}. \textbf{(c)} Transverse arc length $s(x)$ as a function of the coordinate along the constriction for the planar and 3D geometries. Although the projected width narrows from 400 to 200\,nm within the constriction, the 3D architecture preserves a transverse arc length of approximately 400\,nm throughout the structure.}
    \label{fig:fig_1}
\end{figure}

To isolate the influence of 3D shaping on superconducting transport, we compare two architectures with identical in-plane footprints: a planar constriction and a 3D curved constriction [Fig.~\ref{fig:fig_1}(a)]. The planar constriction consists of a superconducting strip measuring 1.2~$\mu$m in length and 400~nm in width, incorporating a central narrowing of a length of 400~nm and a width of 200~nm. The 3D constriction is designed to have the same projection onto the $x$--$y$ plane as the planar geometry while extending into the out-of-plane ($z$) direction. Its profile is engineered such that the arc length of each transverse cross-section remains at least 400~nm [Fig.~\ref{fig:fig_1}(b,c)], thereby preserving the available current-carrying path despite the lateral narrowing. Throughout this work, the membrane thickness is fixed at $d=50$~nm. The material parameters correspond to the direct-write Nb--C-FIBID superconductor \cite{por19acs}, with a zero-temperature coherence length $\xi(0)\sim6$~nm and a penetration depth $\lambda(0)\sim1~\mu$m. The complete set of parameters used in the simulations is summarized in Table~\ref{tab:parameters}.

The system response is evaluated under a dc transport current, $I_\mathrm{tr}$, and an external magnetic field, $\mathbf{B}$, applied along one of the Cartesian axes ($x$, $y$, or $z$). The magnetic field satisfies $B \ll B_\mathrm{c2}=\Phi_0/(2\pi\xi^2)$, where $\Phi_0\approx2.068\times10^{-15}$~Wb is the magnetic flux quantum and $B_\mathrm{c2}$ the upper critical field. Under these conditions, tangential field components contribute only weakly to orbital pair breaking and are therefore neglected \cite{Bra95rpp}. Accordingly, the superconducting response is governed by the local normal projection of the applied magnetic field, which varies across the curved 3D geometry.

Numerical simulations are performed by solving the TDGL equation using the finite element method, as described in Appendix. The TDGL equation is solved under the isothermal approximation, neglecting electrothermal feedback and local temperature variations induced by vortex motion. The validity of this assumption is assessed a posteriori through estimates of the dissipated Joule power. Unless otherwise stated, the results are obtained within the thin-film approximation, which enables efficient exploration of the parameter space. Selected cases are further analyzed using full 3D simulations to verify the vortex-nucleation mechanisms discussed in what follows. As a result, the 3D architecture acquires magnetic-field sensitivities and transport functionalities that are absent in its planar counterpart. This contrast enables us to directly identify how 3D shaping modifies current redistribution, screening-current topology, and vortex nucleation.

\begin{figure*}[t!]
    \centering
    \includegraphics[width=17.8cm]{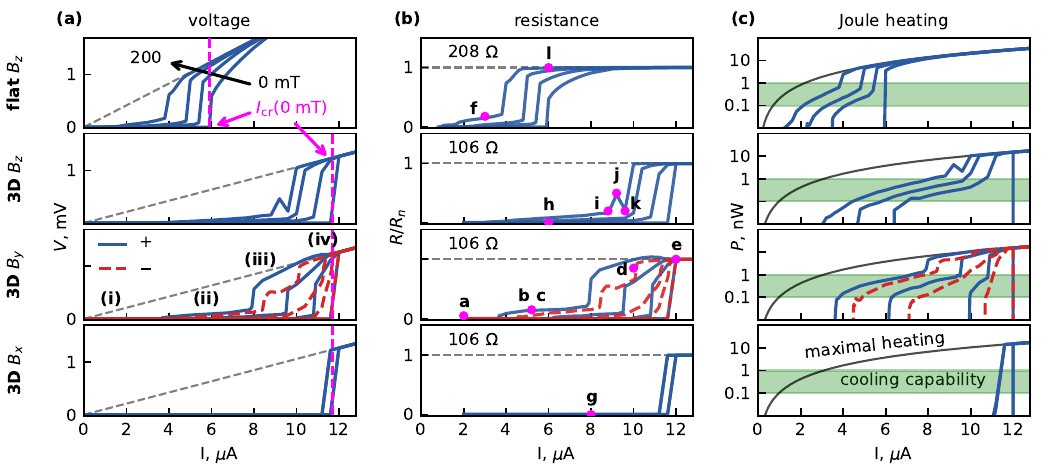}
    \caption{\textbf{(a)} Current--voltage characteristics, \textbf{(b)} resistance as a function of transport current, and \textbf{(c)} Joule heating power for the planar constriction in an out-of-plane magnetic field and for the 3D constriction for a series of values of the magnetic field applied along the three principal directions. The results are shown for $\left|\mathbf{B}\right|=0$, $50$, $100$, and $200$\,mT. For the 3D constriction with a magnetic field oriented along the $y$ axis, opposite current polarities are distinguished by blue solid and red dashed curves. No non-reciprocal response is observed for the other magnetic-field orientations. The Meissner state (i), flux-flow regime (ii), normal domain formation (iii), and fully normal state (iv) are indicated in Fig.~\ref{fig:fig_2}(a). The superconducting order-parameter distributions corresponding to the labeled points \textbf{a}--\textbf{l} in panel (b) are presented in Fig.~\ref{fig:fig_5}(b). The green shaded region in panel (c) indicates the regime where the Joule heating power exceeds the cooling capacity of the substrate, causing the structure to transition into the fully normal state. The black curve labeled ``maximal heating'' represents the Joule heating power expected if the entire nanoarchitecture were in the normal state.}
    \label{fig:fig_2}
\end{figure*}

\section{Results}

\subsection{Geometry-induced current decrowding}

Figure~\ref{fig:fig_2}(a) shows the current--voltage characteristics (CVC) of the planar and 3D constrictions. At zero magnetic field, the 3D architecture exhibits a critical current of approximately $12~\mu$A, nearly twice that of the planar constriction. This enhancement is the first manifestation of the geometry-induced decrowding effect: the additional current-carrying path provided by the 3D geometry suppresses the local current-density enhancement characteristic of planar constrictions, delaying the onset of depairing and vortex nucleation.

The resistive response of the constrictions depends strongly on the orientation of the applied magnetic field. Within the thin-film approximation, the planar constriction responds primarily to the out-of-plane field component, $B_z$  \cite{Gen66boo}, whereas the 3D geometry couples to all field orientations through its curved surface \cite{bog25cpc}. As shown below, this additional degree of freedom gives rise to transport phenomena absent in the planar counterpart.

For both architectures, four transport regimes can be identified in Fig.~\ref{fig:fig_2}(a): (i) a dissipationless Meissner state ($V=0$), (ii) a flux-flow regime in which the voltage increases quasi-linearly with applied current, (iii) an abrupt voltage jump marking the transition to the highly resistive state, and (iv) a fully normal state with an Ohmic response. In the 3D constriction at $B_z=200$~mT, an additional short voltage jump (point j) indicates a reconfiguration of the vortex-flow pattern and normal-domain structure, as discussed below.

Figure~\ref{fig:fig_2}(b) shows the normalized resistance, $R/R_\mathrm{n}$, as a function of the transport current. The simulated normal-state resistances ($208~\Omega$ and $106~\Omega$ for the planar and 3D constrictions, respectively) are slightly lower than the corresponding values estimated from an ideal rectangular geometry ($228~\Omega$ and $114~\Omega$). This deviation originates from the smooth transitions between the wide leads and the constriction region, which modify the current distribution and reduce the effective resistance compared with the rectangular geometry.

\subsection{Superconducting diode effect}
The CVCs for the in-plane transverse field ($B_y$) geometry in Fig.\,\ref{fig:fig_2}(b) exhibit a non-reciprocal critical current, establishing a route toward geometry-engineered superconducting diode functionality.
To quantify the non-reciprocal transport response, we define the diode efficiency as
\begin{equation}
\eta = \left|\frac{
|I_\mathrm{cr}^{+}| - |I_\mathrm{cr}^{-}|
}{
|I_\mathrm{cr}^{+}| + |I_\mathrm{cr}^{-}|
}\right|,
\end{equation}
where $I_\mathrm{cr}^{\pm}$ denotes the critical current for positive ($+$) and negative ($-$) current polarities. The critical current is determined using a voltage criterion of 5~$\mu$V. This criterion is chosen to ensure that the measured signal exceeds the numerical noise level and corresponds to a well-established resistive state associated with vortex nucleation and motion.

\begin{figure}[t!]
    \centering
    \includegraphics[width=8.3cm]{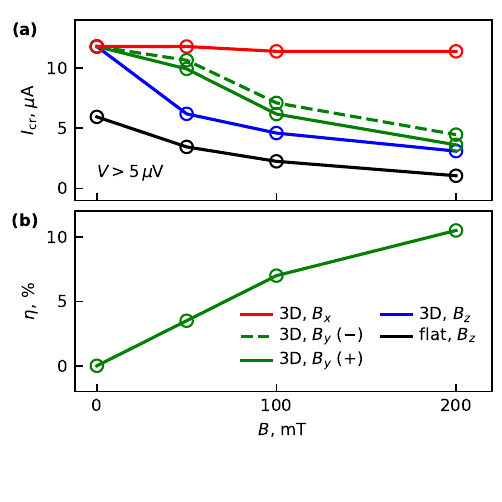}
    \caption{Magnetic-field dependence of the critical current \textbf{(a)} and the  diode efficiency \textbf{(b)} for the 3D constriction under a $y$-oriented magnetic field. 
    }
    \label{fig:fig_3}
\end{figure}

The magnetic-field dependence of $I_\mathrm{cr}$ reveals the impact of the 3D geometry on vortex-mediated dissipation, see Fig.\,\ref{fig:fig_3}. For an out-of-plane magnetic field ($B_z$), the 3D constriction exhibits a critical current nearly twice that of the planar reference structure over the entire field range, consistent with the 3D current decrowding effect. In contrast, the critical current remains largely insensitive to $B_x$, whereas an in-plane transverse field ($B_y$) produces a pronounced splitting between positive and negative current polarities. The resulting diode efficiency, $\eta$, is summarized in Fig.~\ref{fig:fig_3}(b) for the $B_y$ orientation. The non-reciprocal response increases with increasing field magnitude, reaching $\eta \approx 10\%$ at $200$~mT. In this way, by engineering the constriction geometry in three dimensions, the current distribution and vortex-entry landscape become intrinsically asymmetric, enabling tunable non-reciprocal transport controlled by an in-plane magnetic field---a response that is absent in the planar geometry.

\section{Discussion}
\subsection{Screening currents}
\begin{figure}[t!]
    \centering
    \includegraphics[width=8.5cm]{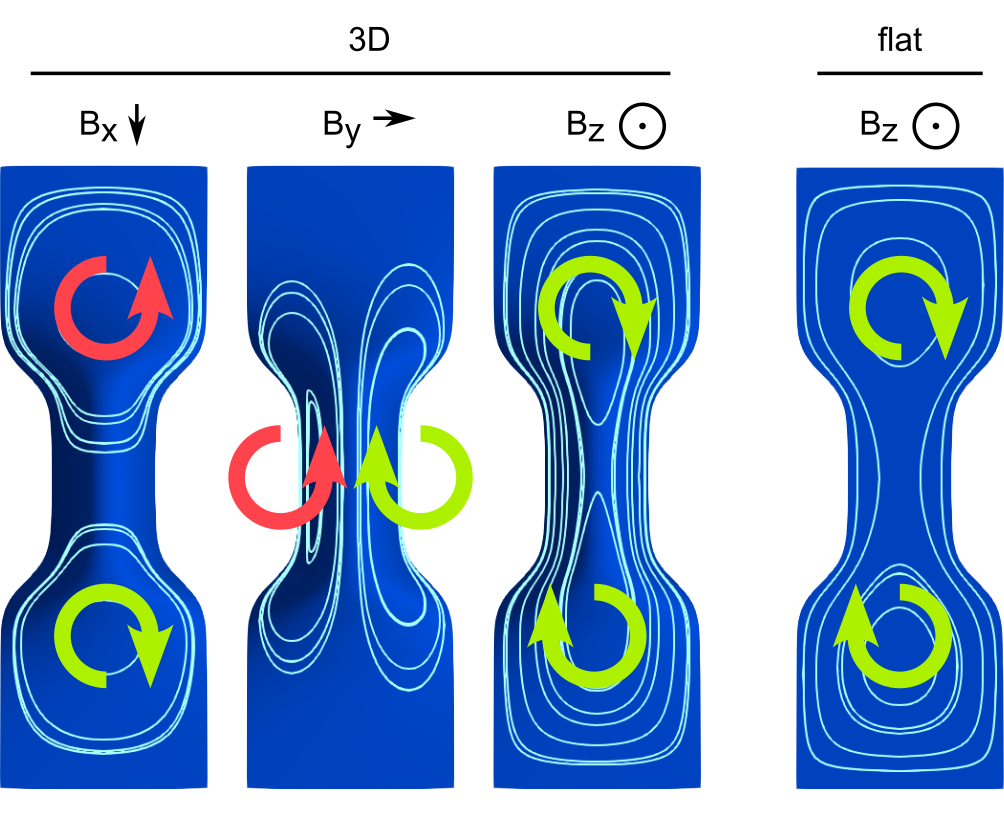}
    \caption{Superconducting screening-current patterns induced by a weak magnetic field in the 3D and planar constrictions. Red and green arrows denote opposite current vorticities. Due to its planar geometry, the flat constriction is sensitive mainly to the out-of-plane field component, while the 3D constriction supports screening-current textures for all three magnetic-field orientations. In particular, an in-plane field produces a pair of counter-rotating screening-current loops.}
    \label{fig:fig_4}
\end{figure}

Our analysis suggests that the screening-current topology is highly sensitive to both the magnetic-field orientation and the 3D geometry of the constriction, see Fig.~\ref{fig:fig_4}. For an out-of-plane field ($B_z$), both planar and 3D structures exhibit a dominant screening loop circulating around the entire perimeter. In the wider lead regions, additional secondary loops emerge with the same clockwise circulation direction. In contrast, the in-plane field $B_x$ generates two spatially separated screening loops with opposite circulation directions, primarily localized within the wide sections of the structure. For the transverse in-plane field $B_y$, the 3D constriction exhibits oppositely circulating screening-current loops on the two sides of the constriction, arising from the opposite normal components of the applied field projected onto the curved surfaces. In this way, 3D geometry couples field orientation to screening-current topology.

\begin{figure*}[t!]
    \centering
    \includegraphics[width=17.8cm]{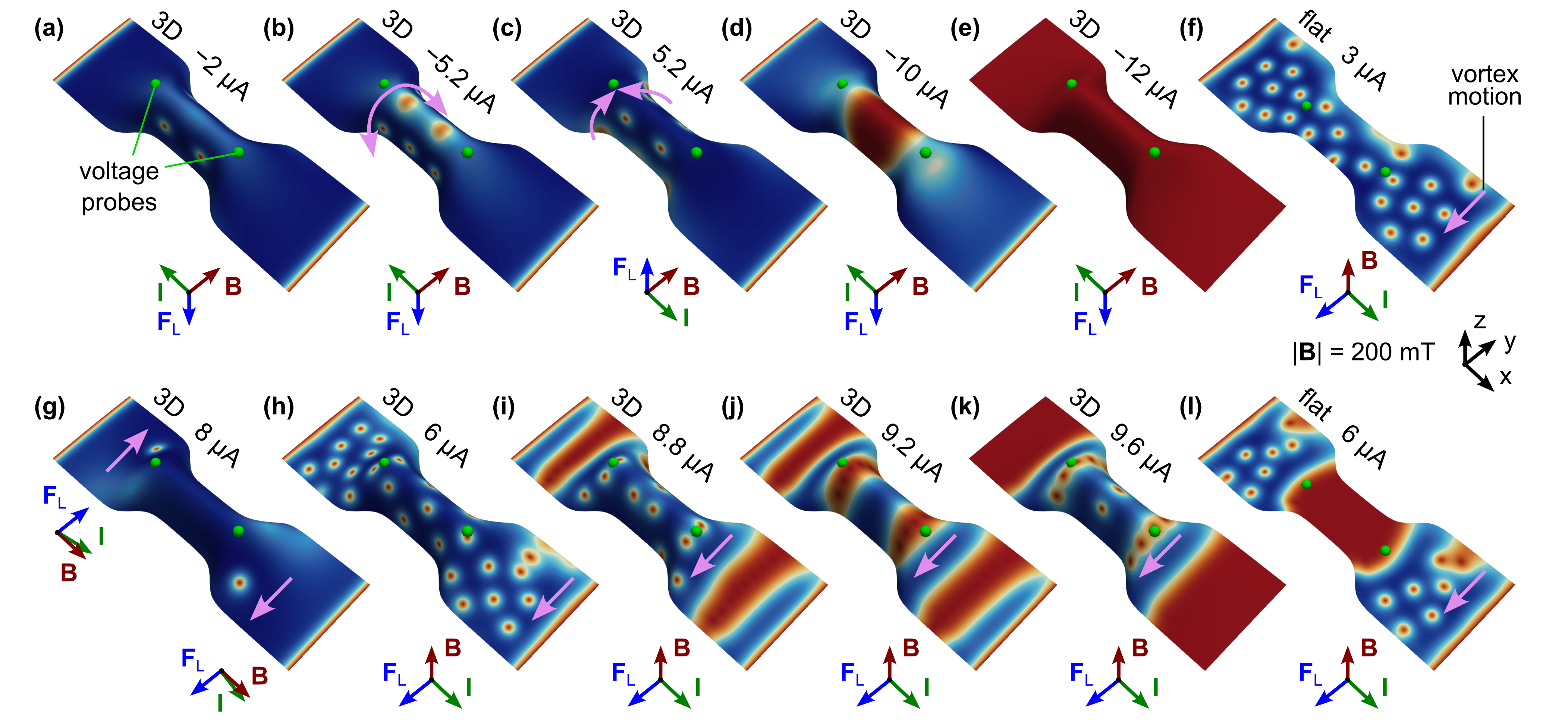}
    \caption{Spatial distributions of the normalized superconducting order parameter magnitude $\left|\psi\right|/\psi_\infty$ for the 3D \textbf{(a--e, g--k)} and flat \textbf{(f, l)} constrictions under different magnetic field orientations and transport current values. Panels (\textbf{a} to \textbf{l}) correspond to the points labeled in Fig.~\ref{fig:fig_2}. Purple arrows denote the direction of vortex motion.}
    \label{fig:fig_5}
\end{figure*}

The screening currents, in conjunction with the transport current, determine the appearance of a non-reciprocal response. The total current-density distribution $\mathbf{j}_\mathrm{tr} + \mathbf{j}_\mathrm{scr}$ must break the symmetry under transport-current polarity reversal ($I\rightarrow -I$) for such a response to occur. For the $B_x$ and $B_z$ field orientations, the two reflection symmetry axes of the geometry ensure that $\mathbf{j}$ remains invariant under this reversal, suppressing any non-reciprocal response. In contrast, the 3D-induced screening-current loops generated by the $B_y$ field direction break this symmetry. Accordingly, depending on the polarity of $I$, the screening currents either reinforce the transport current at the constriction edges while opposing it near the apex, or produce the opposite redistribution. In the former case, the local current density at the edges rapidly approaches the depairing limit, triggering vortex nucleation at reduced transport currents. For the opposite polarity, the screening and transport currents partially cancel at the edges and instead enhance the current concentration near the apex. As a result, vortex entry from the edges is suppressed, and, as shown in the following section, vortices nucleate internally through the apex region of the constriction. This polarity-dependent crossover between edge and internal vortex nucleation gives rise to the observed non-reciprocal critical current, representing a geometry-driven superconducting diode effect analogous to that in wires with triangular cross-sections \cite{mol25nat}.

\subsection{Vortex dynamics}

The distinct transport regimes identified in the CVCs are resolved by examining the superconducting order-parameter distribution, $|\psi|$ [Fig.~\ref{fig:fig_5}]. For the planar constriction under a $B_z$ field, the transition from the flux-flow regime at $3~\mu$A (f) to the resistive normal state at $6~\mu$A (l) occurs when the local current density in the constriction reaches the depairing limit.

By contrast, the 3D geometry exhibits a fundamentally different, multi-stage transition under the same field orientation (h--k). At $6~\mu$A (h), the 3D constriction remains in a stable flux-flow state, whereas the planar reference membrane has already transitioned to the normal state, reflecting the enhanced current-carrying capability provided by 3D current decrowding. Increasing the current to $8.8~\mu$A (i) leads to the formation of vortex rivers \cite{Emb17nac,Bez22prb} in the regions adjacent to the constriction, while the constriction itself remains in a dense flux-flow state. At $9.2~\mu$A (j), additional vortex channels develop within the constriction, producing the localized voltage increase observed in the CVC. At $9.6~\mu$A (k), the adjacent regions become normal while vortex transport is re-established through the constriction. Although these intermediate states are stable within the isothermal TDGL model, in experiments the associated Joule dissipation would likely induce thermal feedback and drive the system into the normal state before all of these transitions become observable. Further increase of the transport current ultimately suppresses superconductivity throughout the entire structure.

To assess the validity of the isothermal approximation, we estimate the dissipated Joule power, $P=IV$ [Fig.~\ref{fig:fig_2}(c)]. Although local current inhomogeneities and voltage fluctuations may introduce deviations, this estimate provides a first-order estimate of the generated heat. Assuming a thermal contact conductance in the range of $10^3$--$10^4~\mathrm{Wm^{-2}K^{-1}}$ \cite{Bez19prb} and a temperature increase of $\Delta T=0.3$~K, corresponding to a fraction of the available margin below the critical temperature, the substrate cooling capacity is estimated to be approximately $0.1$--$1$~nW. We therefore define this range as an approximate thermal stability limit, indicated by the green shaded region in Fig.~\ref{fig:fig_2}(c). Dissipation below this level can be efficiently removed by the substrate, whereas approaching this limit may activate thermal feedback, destabilizing the superconducting state and reducing the experimentally observed critical current.

The in-plane field orientations reveal an additional functionality enabled by the 3D geometry. For $B_x$, at $8~\mu$A (g), the constriction supports two counter-propagating vortex structures, originating from the reversal of the field projection relative to the local surface normal on opposite slopes of the curved membrane.
The most pronounced geometric effects occur for the transverse field $B_y$, where the non-reciprocal response emerges. At low current (a, $-2~\mu$A), vortices remain pinned within the structure. At $\pm5.2~\mu$A (b,c), the current polarity selects between two distinct vortex-entry pathways: vortices enter conventionally from the edges for one polarity, whereas for the opposite polarity they nucleate internally near the apex of the 3D constriction. At $-10~\mu$A (d), superconductivity is suppressed locally within the constriction while the surrounding regions remain superconducting. Complete suppression of superconductivity occurs only at higher current ($-12~\mu$A, e). Together with the critical-current modulation shown in Fig.~\ref{fig:fig_3}(a), these results demonstrate that the $B_y$ field acts as a geometric control parameter, selectively tuning the effective superconducting width of the 3D constriction while leaving the planar regions largely unaffected. This spatially selective control of vortex dynamics represents a unique functionality of 3D superconducting architectures beyond conventional 2D fluxonic devices.

\subsection{Internal vortex nucleation}

The observation of internal vortex nucleation in our 2D simulations raises a question regarding topological conservation. Namely, since vortices are topological excitations characterized by quantized phase winding, changes in the vortex number require either boundary crossing or local suppression of the order-parameter amplitude. Within a purely 2D description of a curved surface, internal vortex nucleation can be interpreted as the simultaneous creation of a vortex--antivortex pair with opposite winding numbers, preserving the net topological charge. However, the full 3D FEM simulations reveal a distinct nucleation pathway enabled by the finite curvature and thickness of the structure.

\begin{figure}[t!]
    \centering
    \includegraphics[width=8.5cm]{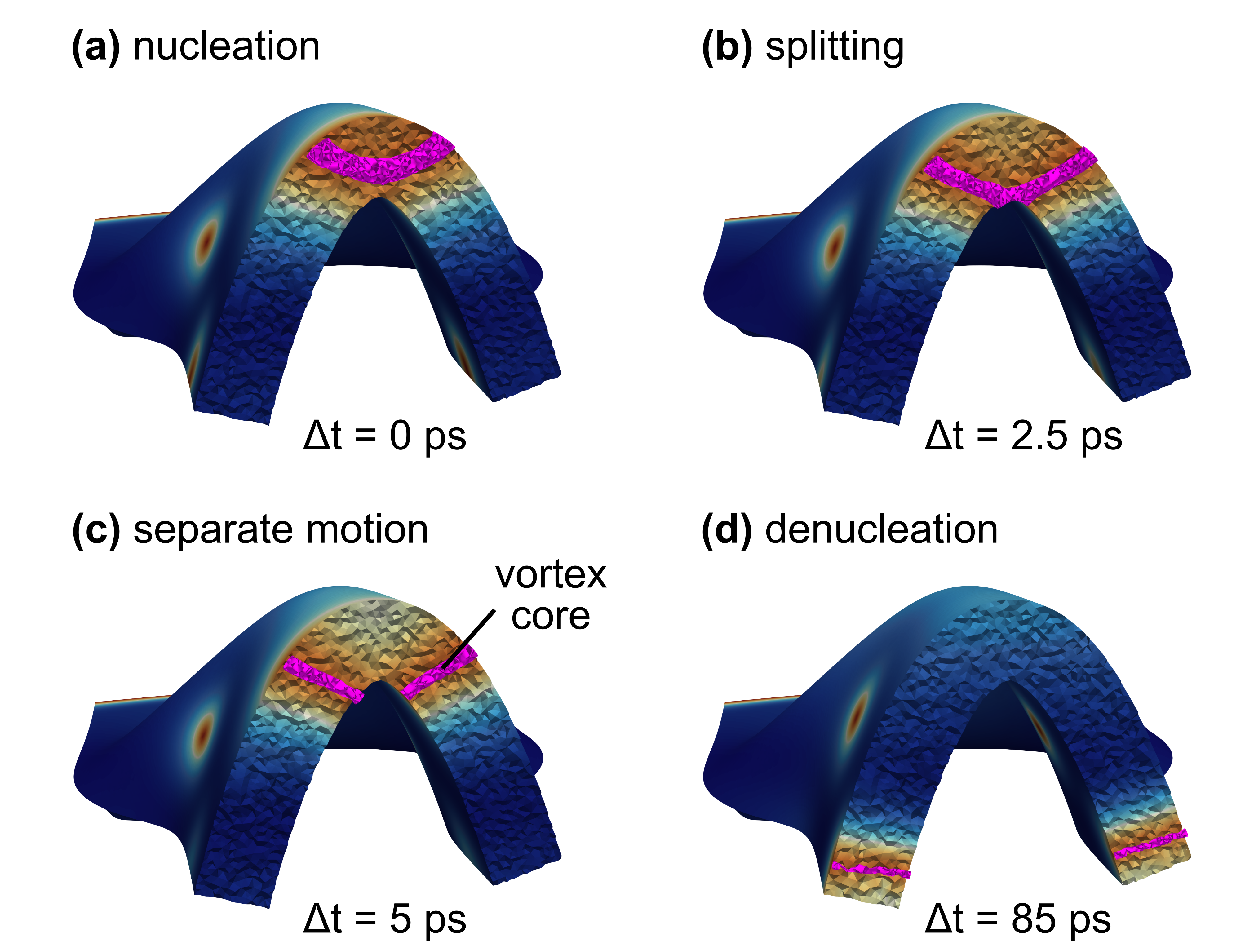}
    \caption{\textbf{(a--d)} Stages of vortex nucleation at the apex of the 3D constriction under a $y$-oriented magnetic field, followed by its splitting into two vortex filaments. Simulations are performed using a full 3D finite-element mesh at a magnetic field of 200~mT and a transport current of 6~$\mu$A. The vortex core is highlighted in magenta.}
    \label{fig:fig_6}
\end{figure}

To address this question, we investigate the nucleation process under a transverse field of $B_y=200$~mT at a transport current of $6~\mu$A using a full 3D FEM mesh. As shown in Fig.~\ref{fig:fig_6}, vortex nucleation begins with the penetration of a single vortex line through the apex region of the curved membrane. The initial vortex segment forms a horizontally oriented vortex core within the membrane plane [Fig.~\ref{fig:fig_6}(a)]. As this vortex line evolves and propagates through the membrane thickness [Fig.~\ref{fig:fig_6}(b)], it undergoes a 3D reconfiguration (splitting), resulting in two vertically oriented vortex filaments extending normal to the membrane surface [Fig.~\ref{fig:fig_6}(c)].

The apparent ``antivortex'' character in the 2D representation therefore originates from the projection of a 3D vortex configuration onto the curved surface, rather than from the creation of a true antivortex. Although the two resulting vortex segments retain the same orientation with respect to the applied magnetic field, their orientations relative to the local surface normal are opposite due to the surface curvature. Consequently, the projected 2D description assigns opposite winding numbers to the two vortex segments, giving rise to the appearance of a vortex--antivortex pair. The two surface-normal vortex filaments subsequently separate [Fig.~\ref{fig:fig_6}(d)] under the action of the Lorentz force and eventually exit through opposite edges of the constriction. Unlike the Berezinskii–Kosterlitz–Thouless (BKT) mechanism, where thermally excited vortex–antivortex pairs subsequently unbind at the transition \cite{Kos16rpp}, the vortex–``antivortex'' pairs considered here are generated deterministically by current-driven topological constraints imposed by the closed superconducting geometry.

Although the 3D vortex-line reconfiguration extends beyond the assumptions of the 2D model---where the order parameter is taken to be uniform across the film thickness---the 2D simulations still reproduce the essential features of the CVCs, including the polarity-dependent internal nucleation process. The full 3D analysis confirms that the observed non-reciprocity arises from the intrinsic coupling between superconducting topology, 3D geometry, and magnetic-field orientation. These results establish 3D nanostructuring as a powerful route toward controlling vortex dynamics and realizing advanced superconducting electronic functionalities.

Finally, we comment on possible experimental realizations of the structures considered here. The required 3D geometries could, in principle, be fabricated either by direct-write deposition of superconducting structures using FIBID/FEBID~\cite{por19acs,cor19nan} or by site-selective chemical vapor deposition onto preformed 3D scaffolds~\cite{Por23acs}. Since our simulations employ material parameters representative of Nb--C fabricated by FIBID~\cite{por19acs}, this material provides a natural candidate platform for experimentally testing the predictions of this work. In addition, superconducting W--C nanostructures fabricated by He$^+$-FIBID~\cite{cor19nan}, which exhibit comparable superconducting length scales and can be directly realized in complex 3D geometries, provide another promising platform for testing the predicted geometry-induced current decrowding and magnetic-field response. Finally, the flat and 3D constrictions could be integrated as active elements into otherwise identical coplanar-waveguide center conductors, with tapered transitions providing coupling to broadband microwave lines \cite{Dob20pra,Dob15met}. This configuration would enable combined dc and microwave excitation and a direct frequency-resolved comparison of geometry-dependent current redistribution, microwave losses, and vortex dynamics, thereby extending the predicted current-decrowding effect from dc transport into the GHz regime.

\section{Conclusions}

We have numerically investigated the current--voltage characteristics and vortex dynamics of superconducting constrictions extended into the third dimension. By comparing 3D and planar constrictions, we show that the 3D geometry can approximately double the critical current through a geometry-induced current-decrowding effect. This enhancement originates from the preservation of the effective current-carrying cross-section afforded by the curved geometry, which suppresses the local current-density enhancement characteristic of planar constrictions. Unlike conventional planar optimization strategies, this approach exploits the third dimension as an additional degree of freedom for controlling current flow.

The 3D constriction also exhibits a pronounced sensitivity to the orientation of the applied magnetic field. Whereas, at moderate field magnitudes, the flat constriction responds primarily to the out-of-plane component, $B_z$, the 3D geometry couples additionally to in-plane field components. In particular, a transverse field $B_y$ enables external tuning of the critical current while leaving the adjacent planar leads largely unaffected. This geometry-enabled spatial selectivity provides local control over superconducting transport that is inaccessible in conventional planar fluxonic devices.

Our simulations uncovered a distinct mechanism of vortex nucleation at the apex of the 3D constriction. We have demonstrated that under a transverse $B_y$ field, vortices do not enter through the edges but instead nucleate at the apex of the 3D constriction in pairs with opposite windings. 3D simulations have revealed that at the moment of such pair nucleation, the vortex core represents a single thread penetrating the thickness of the membrane. This transition from edge to internal nucleation is the fundamental origin of the observed symmetry breaking and the resulting non-reciprocal transport, with a diode efficiency $\eta$ reaching up to $10\%$.

In all, our findings suggest that extending superconducting membranes into the third dimension enables new architectures featuring field selectivity, non-reciprocal transport, and geometry-induced current-decrowding effect. Our findings thus open a new design space for superconducting devices operating close to the depairing limit, with potential implications for single-photon detectors, superconducting bolometers, and Josephson devices. 

\acknowledgments
I.B. acknowledges support from the Deutsche Forschungsgemeinschaft (DFG, German Research Foundation) through Grant No. 583926508 (SyMPaThY) and the use of the CryoMind and CryoFryer simulation workstations at CryoQuant/TU Braunschweig. O.D. acknowledges support from the DFG through Grant No. 577881064 (Super3DMag). This research is based on work from COST Actions CA21144 (SuperQuMap) and CA23134 (Polytopo), supported by the European Cooperation in Science and Technology.

\section*{DATA AVAILABILITY STATEMENT}
The data supporting the findings of this study are openly available\,\cite{Bog26DSDec}.

\section*{Appendix}\label{app:A}
\subsection*{Physical model and approximations}

The simulations are performed within the thin-film approximation ($d\lesssim4\xi(T)$), in which the superconducting order parameter is assumed to be approximately uniform across the membrane thickness. The superconducting response is calculated using the TDGL formalism on a 2D manifold representing the membrane surface. Selected calculations requiring resolution of vortex dynamics across the thickness are additionally performed using the full 3D formulation.

The membrane thickness is fixed at $d=50$~nm throughout this work. The superconducting parameters are representative of direct-write Nb--C \cite{por19acs}; the complete set of material parameters is summarized in Table~\ref{tab:parameters}. The magnetic fields considered in this work satisfy $B \ll B_\mathrm{c2}=\Phi_0/(2\pi\xi^2)$,
where $\Phi_0$ is the magnetic flux quantum and $B_\mathrm{c2}$ is the upper critical field. Under these conditions, the superconducting response is governed primarily by the magnetic-field component normal to the membrane surface \cite{Gen66boo}.

\begin{table}[b!]
\centering
\caption{Superconducting material parameters used in the TDGL simulations for Nb--C.}
\label{tab:parameters}
\renewcommand{\arraystretch}{1.3}
\begin{tabular}{lcc}
\toprule
\textbf{Parameter} & \textbf{Expression} & \textbf{Value}\\
\midrule
Critical temperature 
& $T_\mathrm{c}$ 
& 5.6~K \\

Coherence length at 0 K 
& $\xi(0)$
& 6~nm \\

Coherence length at 5 K
& $\xi(T)=\frac{\xi(0)}{\sqrt{1-T/T_\mathrm{c}}}$
& 18.3~nm \\

Penetration depth at 0 K 
& $\lambda(0)$
& 1~$\mu$m \\

Penetration depth at 5 K
& $\lambda(T)=\frac{\lambda(0)}{\sqrt{1-T/T_\mathrm{c}}}$
& 3.1~$\mu$m \\

GL parameter
& $\kappa=\lambda/\xi$
& 167 \\

Pearl length at 5 K
& $\Lambda=2\lambda^2/d$
& 373~$\mu$m \\

Normal resistivity
& $\rho_\mathrm{n}$
& 570~$\mu\Omega$ cm \\

Relaxation time at 5 K
& $\tau=\pi\hbar/[8k_\mathrm{B}(T_\mathrm{c}-T)]$
& 5.1~ps \\

Critical current at 5 K
& $I_\mathrm{c}=\frac{\hbar wd}{3\sqrt3 e\mu_0\lambda^2\xi}$
& 12~$\mu$A\footnotemark[1] \\

Upper critical field at 5 K
& $B_\mathrm{c2}=\Phi_0/\left(2\pi\xi^2\right)$
& 1.0~T \\

\bottomrule
\end{tabular}

\footnotetext[1]{For a film cross-section of $wd=400\times50$~nm$^2$.}
\end{table}

Self-induced magnetic fields generated by the transport current are neglected. This approximation is justified by the extreme type-II character of Nb--C ($\kappa \approx 167$) and the large Pearl length \cite{Pea66jap}, $\Lambda= 2\lambda^2/d\approx384~\mu\mathrm{m}$, which exceeds the characteristic lateral dimensions of the constriction by more than three orders of magnitude. Consequently, magnetic-field penetration is essentially complete and current-generated stray fields remain negligible compared with both the applied field and the characteristic fields associated with vortex nucleation.

Throughout this work, we adopt the isothermal approximation. Electrothermal feedback and local heating associated with vortex motion are not included explicitly. The validity of this approximation is assessed a posteriori through estimates of the dissipated Joule power discussed in the Results section. Finally, the model describes an ideal superconducting structure and neglects edge roughness, material inhomogeneity, and thermal fluctuations. These effects may influence vortex nucleation and modify the quantitative values of the critical currents \cite{Zot14sst,Bud22pra}, but are not expected to alter the geometric mechanisms discussed here.

\subsection*{Time-dependent Ginzburg--Landau formalism}

The numerical simulations were performed by solving the TDGL equation

\begin{equation}
    \tau \left(
        \partial_t + i \frac{2 e}{\hbar} \phi
    \right) \psi
    =
      \xi^2 \left(
        \nabla - i \frac{2 e}{\hbar} \mathbf{A}
    \right)^2 \psi
    + (1 - |\psi|^2) \psi
\end{equation}
coupled to the Poisson equation
\begin{equation}
    \nabla \cdot( \sigma_\mathrm{n} \nabla \phi )
    =
    \nabla \cdot \mathbf{j}_\mathrm{sc},
\end{equation}
where the superconducting current density is
\begin{equation}
    \mathbf{j}_\mathrm{sc}
     = \frac{\hbar}{2 \mu_0 e \lambda^2}
     \Im\left\{
        \psi^*
        \left(
        \nabla - i \frac{2e}{\hbar}\mathbf{A}
        \right)\psi
    \right\}.
\end{equation}

Here, $\psi$ denotes the superconducting order parameter normalized to unity in equilibrium, $\mathbf{A}$ is the magnetic vector potential, $\mathbf{B}=\nabla\times\mathbf{A}$ is the magnetic field, $\phi$ is the electric scalar potential, and $\sigma_\mathrm{n}=1/\rho_\mathrm{n}$ is the normal-state conductivity.

The free-boundary conditions are
\begin{equation}
\left.
\begin{aligned}
\mathbf{n}\cdot
\left(
\nabla-i\frac{2e}{\hbar}\mathbf{A}
\right)\psi &=0,\\
\mathbf{n}\cdot\nabla\phi &=0
\end{aligned}
\right\}
\qquad \text{on } \Gamma_\mathrm{free},
\end{equation}
whereas ideal metallic contacts are described by
\begin{equation}
\left.
\begin{aligned}
\psi &=0,\\
\mathbf{n}\cdot\nabla\phi
&=
-\rho_\mathrm{n}j_\mathrm{tr}
\end{aligned}
\right\}
\qquad \text{on } \Gamma_\mathrm{c}.
\end{equation}

\subsection*{Geometry definition}

The geometry of the 3D constriction is defined by the analytical parametrization
\begin{align}
    w(\zeta) &= \frac{1}{4}\left(4 - \tanh\left(\frac{\zeta - 400}{u}\right) + \tanh\left(\frac{\zeta - 800}{u}\right)\right), \\
    x &= \zeta, \nonumber\\
    y &= \eta\, w(\zeta), \nonumber\\
    z &= h \left(1 - w(\zeta)^2\right)^{1/4}
         \left(1 - \left(\frac{\eta}{200}\right)^2\right), \nonumber
\end{align}
where $u=50$~nm controls the sharpness of the in-plane constriction and $h=175$~nm defines the out-of-plane height. The coordinates span
$\zeta\in[0,1200]$~nm and $\eta\in[-200,200]$~nm.
The planar constriction is recovered by setting $z=0$.

For the 2D simulations, the fields $\psi$ and $\phi$ are assumed to be approximately uniform across the membrane thickness. The governing equations are therefore projected onto the tangent manifold of the superconducting surface following Refs.~\cite{cos81pra,bog25cpc}.

\subsection*{Numerical implementation}

All equations are formulated in weak form and solved within a finite-element framework using FEniCS \cite{aln15ans}. Time integration is performed using the Crank--Nicolson scheme, while the scalar potential is incorporated through a gauge-invariant link-variable formulation \cite{Kat93prb}. The time step is set to $0.25$~ps and $0.5$~ps for the 2D and 3D simulations, respectively. The gauge transformation $\phi\rightarrow\phi+\mathrm{const}$ introduces a nullspace in the linear system. This degree of freedom is removed using a Lagrange multiplier that constrains the spatial average of $\phi$ to zero.

The 2D meshes are generated using Gmsh \cite{geu09nme}. For the 3D simulations, the surface geometry is first imported into Blender \cite{blender}, where a solidify operation is applied to generate a volumetric body, followed by remeshing in Gmsh. Mesh statistics are summarized in Table~\ref{tab:mesh}.
\begin{table}[h]
    \centering
    \caption{Mesh statistics for the 2D and 3D constriction geometries.}
    \label{tab:mesh}
    \begin{tabular}{lccc}
        \hline
        \textbf{Configuration} &
        \textbf{Nodes} &
        \textbf{Surface cells} &
        \textbf{Volume cells} \\
        \hline
        3D constriction (surface)
        & 24\,358 & 48\,048 & --- \\

        3D constriction (solidified)
        & 377\,513 & --- & 1\,944\,419 \\

        2D constriction (surface)
        & 7\,630 & 14\,842 & --- \\

        2D constriction (solidified)
        & 359\,454 & --- & 1\,897\,712 \\

        \hline
    \end{tabular}
\end{table}

CVCs are obtained using a quasi-steady-state protocol. The system is initialized with
$\phi=0$ and $\psi=1+\mathrm{noise}$ and relaxed for 10~ps. The magnetic field is then ramped linearly over 50~ps, followed by a further 40~ps relaxation period. Subsequently, the transport current is increased linearly over 500~ps, and the system is allowed to evolve for an additional 1~ns. The voltage is evaluated as the time-averaged difference of $\phi$ between two voltage probes over a subsequent 1~ns sampling window \cite{Bog22prb}.

To visualize the screening-current patterns in Fig.~\ref{fig:fig_4}, simulations were performed at a low magnetic-field magnitude (10~mT) in the absence of transport current. The current streamlines were generated in {ParaView} \cite{paraview} by integrating the vector field of current density using a fourth-order Runge--Kutta scheme, with integration paths initiated from randomly distributed seed points on the surface.

\FloatBarrier

\bibliography{main}

\end{document}